\documentclass[aps,twocolumn,superscriptaddress,showpacs]{revtex4}
\usepackage{epsf,latexsym}
\usepackage{times}
\usepackage{graphicx}
\usepackage{dcolumn}
\usepackage{bm}

\newcommand\R{{\mathrm {I\!R}}}
\newcommand\N{{\mathrm {I\!N}}}

\newcommand{\be}{\begin{equation}}
\newcommand{\ee}{\end{equation}}

\def\CC{{\rm\kern.24em \vrule width.04em height1.46ex depth-.07ex
    \kern-.30em C}}
\def\P{{\rm I\kern-.25em P}}

\def\bbbc{{\mathchoice {\setbox0=\hbox{$\displaystyle\rm C$}\hbox{\hbox
to0pt{\kern0.4\wd0\vrule height0.9\ht0\hss}\box0}}
{\setbox0=\hbox{$\textstyle\rm C$}\hbox{\hbox
to0pt{\kern0.4\wd0\vrule height0.9\ht0\hss}\box0}}
{\setbox0=\hbox{$\scriptstyle\rm C$}\hbox{\hbox
to0pt{\kern0.4\wd0\vrule height0.9\ht0\hss}\box0}}
{\setbox0=\hbox{$\scriptscriptstyle\rm C$}\hbox{\hbox
to0pt{\kern0.4\wd0\vrule height0.9\ht0\hss}\box0}}}}

\def\bbbz{{\mathchoice {\hbox{$\sf\textstyle Z\kern-0.4em Z$}}
{\hbox{$\sf\textstyle Z\kern-0.4em Z$}}
{\hbox{$\sf\scriptstyle Z\kern-0.3em Z$}}
{\hbox{$\sf\scriptscriptstyle Z\kern-0.2em Z$}}}}

\begin{document}
\title{Bi-partite mode  entanglement of bosonic condensates  on  tunnelling graphs}
\author{Paolo Zanardi}
\affiliation{Institute for Scientific Interchange (ISI) Foundation, Viale Settimio Severo 65, I-10133 Torino, Italy}

\begin{abstract}
We study  a set of $L$ spatial bosonic modes localized on a graph $\Gamma.$
The particles are allowed to tunnel from  vertex to vertex by hopping along the edges
of $\Gamma.$ 
 We analyze how,  in the  exact many-body eigenstates of the system i.e., Bose-Einstein condensates over
single-particle eigenfunctions, the bi-partite quantum entanglement
of  a graph vertex with respect to the rest of the graph depends on the topology of $\Gamma.$
\end{abstract}
\maketitle

The possibility of exploiting the quantum features of bosonic particles e.g., cold bosonic atoms, 
living on coupled spatial lattices to  the aim of Quantum Information Processing (QIP) \cite{QIP} 
has been recently addressed in the literature \cite{bec_lattice, bec_qubit, ioniza,duan}.
These systems provide also a unique opportunity  to investigate fascinating coherent  phenomena 
e.g., quantum-phase transitions  \cite{bec_mott}.

In this note we shall study a simple problem related to this more general context.
We shall consider a set of $N$ bosonic particles hopping between the $L$ vertices of a {\em graph}
$\Gamma,$ we will assume the on-vertex self-interaction terms to be zero.
The associated elementary quadratic Hamiltonian is exactly solvable
and many-body eigenstates are simply given by Bose-Einstein consensates (BECs) over  single-particle wavefunctions.
This kind of abstract situation could be realized, for istance, in a optical lattice loaded with cold atomic
atoms that can tunnel from  different local traps and with atom self-interactions
somehow switched off \cite{bec_lattice}.

The aim  is to analyze the role of the graph topology in determining, in those many-body eigenstates,
 the bi-partite quantum entanglement of  a  vertex with respect to the rest of the graph vertices.
In particular one can address the issue of bi-partite entanglement in the ground-state of the system   
and how e.g., for QIP purposes to optimize it by graph designing (for a related study see \cite{bur, simon, hines}).

It is worthwhile to stress that in this paper the view of quantum entanglement
in system of indistinguishable particles is the one based on modes advocated in Refs. \cite{zan,gitt,van}  
rather than the complementary  one based on particles \cite{Schli}.

Let us start  by recalling the basic kinematical framework an to lay down the basic notations.
The quantum state-space associated with graph $\Gamma$ is given by the tensor product of $L$
linear oscillator Fock spaces ${\cal H}_\Gamma \cong \otimes_{j\in\Gamma} \mbox{span} \{|n_j\rangle\}_{n_j=0}^\infty.$
Since  we  are mostly interested in 
massive particles e.g., atoms, we will focus on sectors of ${\cal H}_\Gamma$
with definite total particles number 
\begin{equation}
{\cal H}_\Gamma^{(N)}:= \mbox{span}\{ \otimes_{j=1}^L |n_j\rangle\,/\,\sum_{j=1}^Ln_j=N\}.
\label{FockN}
\end{equation} 
Given a state $|\Psi\rangle\in{\cal H}_\Gamma^{(N)}$ we are here  interested to 
the {\em on-site} reduced density matrix;
if $|\Psi\rangle=\sum_{n_1,\ldots,n_L} C(n_1,\ldots,n_L)\,\otimes_{j=1}^L |n_j\rangle$ one has, say for the first
vertex
\begin{eqnarray}
\rho^{(1)}&:=&\mbox{Tr}_{\Gamma-\{i\}} |\Psi\rangle\langle \Psi|=\sum_{m=0}^N \rho^{(1)}_m
|m\rangle\langle m|,\nonumber \\
\rho^{(1)}_m &=&\sum_{n_2,\ldots,n_L}^\prime |C(m,n_2,\ldots,n_L)|^2.
\label{reduced}
\end{eqnarray}
The prime in the above sum simply reminds that the condition $\sum_{j=2}^L=N-m$ must be fulfilled.
The crucial, though obvious thing, to notice here is that the constraint of fixed total
particle-number results in a {\em diagonal} reduced  density matrix, and that such a matrix
can be always seen as an operator over the finite-dimesional space  $\CC^{M}$ with $M\ge N.$
This remark  relieves us to face with the subtleties of entanglement definition in truly 
infinite-dimensional spaces \cite{infini}

Let the hamiltonian be
\begin{equation}
H[A]=-\sum_{i,j=1}^L A_{ij} b_i^\dagger b_j
\label{H}
\end{equation}
where 1) the $b_i$'s are bosonic modes, 2) $A:=(A_{ij})_{ij}\in M_L(Z_2),\,(Z_2=\{0,1\})$
 is an symmetric  matrix. We will consider the case in which $A$ is an {\em adiacency} matrix of a {\em graph}
\cite{SGT}
$\Gamma=(V,E),$ where $V=\{1,\ldots,L\}$  is the set of {\em vertices} and
$E$ is the set of {\em edges}, $(i,j)\in E$ iff $ A_{ij}\neq 0$

By diagonalizing $A$ one gets $H[A]=\sum_{k=1}^L \omega_k B_k^\dagger B_k,$
where $\omega_k$ are the $A$-eigenavalues and $B_k^\dagger=\sum_{j=1}^L  U_{kj} b_j$
are new bosonic modes ($ U\in M_L(\CC)$ is unitary).

Let us now consider a non-degenerate   eigenvalue $\omega_1$ of $A$ 
and a  $N$ particles condensate over it.
If $B_1=(U_{11},U_{12},\ldots,U_{1L})$ denotes the associated eigenvector, one has
\begin{eqnarray}
|B_1^N\rangle &:=&  \frac{1}{\sqrt{N!}} (B_1^\dagger)^N|0\rangle= \frac{1}{\sqrt{N!}}
\sum_{i_1,\ldots,i_N} \prod_{k=1}^N U_{1,i_k} b_{i_k}^\dagger|0\rangle\nonumber \\
&=& \sqrt{\frac{N!}{\prod_{k=1}^L n_k!}}\otimes_{k=1}^L U_{1,k}^{n_k}|n_k\rangle 
\end{eqnarray}
The reduced density matrix associated to the $i$th mode is given by
\begin{equation}
\rho^{(i)}:=\mbox{Tr}_{\Gamma-\{i\}} |B_1^N\rangle\langle B_1^N|=\sum_{m=0}^N \rho^{(i)}_m(B_1)
|m\rangle\langle m|
\end{equation}
where
\begin{eqnarray}
& & \rho^{(i)}_m(B_1)
=\sum_{\{j_n\}\in {\cal S}_N(i,m) }\prod_{n=1}^N|U_{1,j_n}|^2
\nonumber\\
&=&\sum_{\{n_k\}\in \tilde {\cal S}_N(i,m)}
N!\prod_{l=1}^L\frac{1}{n_l!}|U_{1,l}|^{2 n_l},
\label{rho}
\end{eqnarray}
where
\begin{eqnarray}
{\cal S}_N(i,m)&=&\{ (j_n)\in\N_L^N\,/\,\#\{j_n=i\}=m\}
\nonumber \\
\tilde {\cal S}_N(i,m)&:=&\{ (n_l)\in \N_N^L\,/\, \sum_{l=1}^L n_l=N,\,n_i=m\}.
\end{eqnarray}
Now, using the fact that $\sum_{j=1}^L|U_{1j}|^2=1$ is not difficult to see
that one can further rearrange the last expression in Eq.(\ref{rho})
in oder to get 
\begin{equation}
\rho^{(i)}_m(B_1)=\pmatrix{ N\cr m}|U_{1,i}|^{2m}
(1-|U_{1,i}|^{2})^{N-m}
\label{major}
\end{equation}
This  expression is the result we needed.
Clearly Eq. (\ref{major}) has a very simple  meaning:
the probability $p$ of occupying the vertex $i$ ($\Gamma -\{i\}$) in the single-particle
wavefunction $B_1$ is given by $|U_{1i}|^2,$ ($1-|U_{1i}|^2$).
Since the BEC over $B_1$ is the tensor-product of $N$ copies of $B_1$
the probability $\rho^{(i)}_m$ of having $m$-particle on $i$
is given by a binomial distribution $ \pmatrix{ N\cr m} p^m(1-p)^{N-m}.$
This classical argument  works because of the fixed particle-number constraint
forces the vertex reduced density matrix to be diagonal i.e., 
a probability distribution.

From now on we will measure
 entanglement  by the von Neumann  entropy of the reduced density matrix
\begin{eqnarray}
 e_N^{(i)}(B_1)&:=&S(\rho^{(i)}(B_1))=-\mbox{Tr}(\rho^{(i)}\log_2 \rho^{(i)})
\nonumber\\
 &=&-\sum_{m=0}^N \rho^{(i)}_m \log_2 \rho^{(i)}_m 
\label{S}
\end{eqnarray}
By noting  that $B_1$ can be an {\em arbitary} single-particle wavefunction i.e., non
necessarily an $H[A]$ eigenstate, one realizes that Eq. (\ref{S}) defines -- for any given
vertex $i$ of $\Gamma$ -- a positive 
real-valued function over the single-particle space i.e., $e_N^{(i)}\colon \CC^L\rightarrow \R_0^+.$
From expression (\ref{major})  one readily   show that
\begin{itemize}

\item{} The entanglement of a vertex with respect to the others in a BEC
depends only on the {\em square amplitude}, over the considered site, 
 of the single-particle eigenstate we are condensing over.
\end{itemize}

More formally $e_N^{(i)}(W\,x)=e_N^{(i)}(x),\,(\forall x\in\CC^L)$
for unitaries $W$ belonging to the group $U(1)\times U(L-1)$
(phase on the $i$-th component, arbitary unitary mixing of all the other ones).
This invariance is, of course, nothing but the invariance of entanglement
with respect to local transformations.

\begin{figure}[t]
  \includegraphics[height=6cm, angle=-90 ]{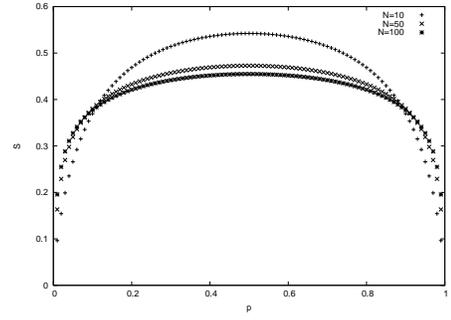}
  \caption{\label{entro0}
Entanglement entropy as a function of $p:=|U_{1i}|^2$ for different particle numbers $N.$
 The value has been normalized to
the maximally available entropy $\log_2(N+1).$ 
}
\end{figure}

\begin{itemize}
\item{} The graph size $L$ does not enter in entanglement properties,
but possibly through the single particle amplitude $|U_{1,i}|.$

\item{}  It is easy to prove that the functions $e_N^{(i)}$'s
have a maximum for $U_{1,i}=1/\sqrt{2}$ (see Fig. (\ref{entro0})
\end{itemize}

\begin{figure}[t]
  \includegraphics[height=6cm, angle=-90 ]{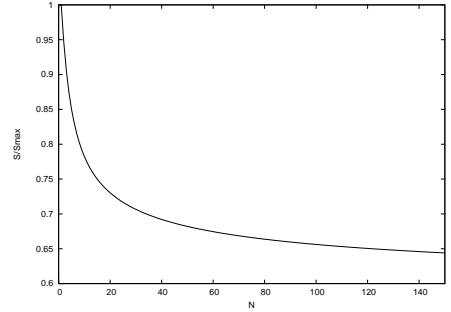}
  \caption{\label{entro}
   Ratio between ${\mbox{max}}\, e_N^{(i)}$  and maximally available entanglement
 as a function of the total particle number.
 }
\end{figure}

The highest achieved value for bi-partite mode entanglement is then given 
by  
\begin{equation}
\mbox{max}\, e_N^{(i)} =N- 2^{-N} \sum_{m=0}^N \pmatrix{ N\cr m}\log_2  \pmatrix{ N\cr m}
\end{equation}
this is a monotonic increasing function of $N,$  but -- due to the well-known properties
of binomial coefficients -- is monotonic decreasing {\em fraction} of the maximun
available entropy $\log_2 (N+1).$ For $N\mapsto \infty$ 
such a fraction seems to attain a finite value. We numerically estimated this asimptotic ratio
to be about $0.57,$ see Fig (\ref{entro}). 

Let us define  ${\cal G}_L$ as the set of all undirected graphs with $L$ vertices
($ |{\cal G}_L|=2^{L(L-1)/2} $),
given $\Gamma\in{\cal G}_L$ one has the  $L$ eigenvectors $B_1(\Gamma),\ldots,B_L(\Gamma)$
of the associated adiacency matrix.
The real-valued functionals   we want to analyse   are given by
\begin{equation}
\Gamma\in{\cal G}_L, \longrightarrow \mbox{max}_k e_N^{(i)}[B_k(\Gamma)]
\label{funct} 
\end{equation} 
One could restrict the problem, by considering just the eigenvector $B_1(\Gamma)$
associated with the {\em largest} eigenvalue of the adiacency matrix of $\Gamma.$
This eigenvector corresponds then (see Eq. (\ref{H}) to the {\em lowest} single-particle
energy and the BEC $|B_1^N\rangle$ is a many-body {\em ground state} of the Hamiltonian (\ref{H}). 
By the Perron-Frobenius theorem \cite{SGT} we know that --for connected $\Gamma$ -- $B_1$ is elementwise positive
and that the associated  eigenvalue is nondegenerate, hence the ground state is unique.  

To exemplify this problem let us consider as 
 $\Gamma$ the complete graph minus the diagonal i.e.,
$A[\Gamma]=\sum_{i,j=1}^L (1-\delta_{i,j})|i\rangle\langle j|.$
By writing this matrix in the following form
$A[\Gamma]= L |X\rangle\langle X|-\openone,\,|X\rangle:= L^{-1/2}\sum_{j=1}^L|j\rangle,$
one immediately realizes that the $A$ spectrum is given by $L-1$ (with eigenvector $|X\rangle$)
and by $0$ with associated the the $L-1$ operators $\tilde b_k,\,(k=1,\ldots,L-1).$
The $N$-particle ground state is therefore provided by putting the $N$
in the $k=0$ bosonic mode associated with $X.$
The ground-state bi-partite entanglement is given by (\ref{major}) and (\ref{S})
with $|U_{1j}|=L^{-1/2}.$

If $\Gamma$ is a {\em regular} graph with connectivity $r$ i.e., all the vertices have $r$ neighbors,
it is fact of elementary spectral graph theory \cite{SGT} that the highest eigenvalue of the $\Gamma$
adiacency matrix is given by $r$
and the associated eigenvector is given by the $0$ Fourier mode $1/\sqrt{L}(1,\ldots,1).$
Therefore for regular graphs maximal bi-partite entanglement is possible 
just for  the dimer i.e., $L=2$.  Notice that for  the more general case of 
one-dimensional  rings  with $L$ (diagonalized by Fourier transformation with cyclic boundary conditions)
the same value  of (\ref{S})  
is achieved for all the vertices in all the BECs 
in single particle eigenstates. This fact stems from translational invariance 
which implies that  all the single-particle eigenfunctions 
have the same vertex square amplitude i.e., $L^{-1}.$

It is interesting to note in passing that the
 {\em bi-partite} graphs $( V=A\cup B,\, (a,b)\in E \Leftrightarrow a\in A $ and $ b\in B )$
the mode entanglement associated with
BEC over the single-particle eigenvalue $E$ is the same as the one associate with eigenvalue $-E.$
One can realize this fact by performing the following canonical transformation in the Fock space
 associated with the $\Gamma$ modes: $c_j\longrightarrow (-1)^{\chi_A(j)} c_j, $ where $\chi_A$ 
denotes the characteristic function of the sub-graph $A.$
One has that $H[A]\longrightarrow H[-A]=-H[A],$ and that the $H[A]$ eigenvectors change their components
over the Fock basis $\otimes_{j=}^L|n_j\rangle$ just by a phase factor $\exp(i\pi\sum_{j\in A} n_j)$.
Then the claim follows straight away from Eq. (\ref{reduced}).
Notice also that this symmetry property implies that for {\em any initial} state $|\Psi\rangle$
(not necesarily an $H[A]$ eigenstate) the on-vertex entanglement dynamics is invariant under time-reversal,
i.e., $S(t)=S(-t),$ and moreover  this result holds
even in presence of local Hubbard-like self interactions \cite{pgpz}

For a general number of vertices $L,$ the natural question is?

{\em What is the graph topology  which  optimize the on-vertex entanglement?}

The answer is not difficult to find out. 
Let $A$ be the adiacency matrix of the ``star'' i.e., just the node $1$ is connected to all the others,
$A_{i,j}=\delta_{i,1}.$ This matrix has two non zero eigenvalues $\epsilon_\pm=\pm \sqrt{L-1}$
corresponding to the single-particle operators 
\begin{equation}
b_\pm:=\frac{1}{\sqrt{2}}(b_1\pm\frac{1}{\sqrt{L-1}}\sum_{j=1}^{L} b_j).
\label{comb}
\end{equation}
The $N$-particle ground state is unique and is given by $|b_+^N\rangle=(b_+)^{\dagger N}/\sqrt{N!}|0\rangle.$
In view of Eqs (\ref{major}) and (\ref{comb}) the functional $e^{1}_N$ is maximized for all $N$
by the star  graph. Physically this means that the star topology optimizes the bi-partite
entanglement in the ground-state BEC. In view of the "monogamy" properties 
of quantum entanglement \cite{QIP} this result looks , in a sense, rather  intuitive.
A naive argument is that 
 the star topology is the one with  maximall connectivty of the vertex $0$
with the  subgraph with  $V=(1,\ldots,L-1)$, this  latter in turn  is totally disconnected
and therefore among its vertices there is small entanglement.

In this brief report we studied the mode  entanglement in Bose-Einstein condensate
over a purely  tunnel-coupled graph. We found an  exact expression for such 
a quantity for arbitray graph and particle number.
We  proved that the star topology maximizes the bi-partite  entanglement
of the spatial mode associated to the star center with the rest of the vertices.
The role of local self-interaction i.e., non-linear terms, as long as the practical relevance
e.g., implementation, QIP protocols, 
of our abstract though simple analysis is  subject of ongoing investigations \cite{pgpz}.

I thank for valuable inputs,   R. Ionicioiu, P. Giorda  and R. Burioni.


\end{document}